\documentclass[print, nolongbibliography,groupedaddress,showpacs,showkeys,amsmath,amssymb,eqsecnum,aps,prd,nofootinbib]{revtex4-2}

\usepackage{enumerate}
\usepackage[]{graphicx}
\usepackage{mathtools}

\usepackage{bm}
\usepackage{verbatim}

\usepackage{hyperref}

\begin{document}

\title{Field redefinition invariant Lagrange multiplier formalism}

\begin{abstract}
In this paper, we propose a field redefinition invariant Lagrange multiplier (LM) formalism in which new ghost-like fields, analogous to Lee--Yang ghosts, are introduced. These ghost fields are required to restore the field redefinition invariance of the standard path integral of the LM theory and, at the same time, to cancel the additional contributions due to the LM fields. We argue that the extra degrees of freedom in the standard LM formalism, coming from the LM fields, should cancel against the degrees of freedom of the ghost fields. Hence, in the field redefinition invariant formalism the doubling of degrees of freedom, associated with the LM fields, is absent.
\end{abstract}



\author{F. T. Brandt}
\email{fbrandt@usp.br}


\author{S. Martins-Filho}
\email{sergiomartinsfilho@usp.br}
\affiliation{Instituto de F\'{\i}sica, Universidade de S\~ao Paulo, S\~ao Paulo, SP 05508-090, Brazil}

\date{\today}
\maketitle

\section{Introduction}\label{section:intro}

In recent years, there has been an increasing interest in the study of higher-derivative theories, especially in the context of gravity, mainly due to the application of such models in cosmology (for a review we refer to Refs. \cite{Clifton:2011jh, Woodard:2006nt, Shankaranarayanan:2022wbx}). Other examples of higher-derivatives theories include noncommutative theories \cite{Douglas:2001ba, Szabo:2001kg} and the Pais-Uhlenbeck oscillator \cite{Pais:1950za, Smilga:2008pr}. The higher-derivative gravity models were conceived, a long time ago, to obtain theories that could tame ultraviolet divergences that originally arise in the quantization of the Einstein-Hilbert action \cite{thooft:1974, Goroff:1985th, vandeVen:1991gw, thirring:1950, Utiyama:1962sn}. Based on this idea, Stelle \cite{Stelle:1976gc} obtained a renormalizable quantum gravity theory introducing quadratic terms in the Einstein-Hilbert action. 

Recently in a series of papers \cite{McKeon:1992rq, Brandt:2018lbe, Brandt:2019ymg, Brandt:2020gms, McKeon:2021qhv, Brandt:2021nev} a simpler alternative has been proposed that restricts quantum effects to one-loop order and has General Relativity as the classical limit.  This is done by using Lagrange Multipliers (LM) fields to restrict the quantum path integral to field configurations that satisfy the Euler-Lagrange equations. The \emph{standard LM theory}, as we shall call it, yields twice the usual one-loop contributions, while the tree-level effects are kept unaltered. The degrees of freedom are also doubled. 
Although it is clear that the doubling of one-loop contributions and degrees of freedom is caused by the presence of the LM fields; its interpretation is, however, still open. 

On the other hand, these extra degrees of freedom appear to be associated with the propagation of ghosts (the kinetic term of the LM fields has the wrong sign) so that the additional one-loop contributions in the standard LM theory may be considered unphysical \cite{McKeon:1992rq}. 
This is also a general downside of non-degenerate higher-derivatives theories, which include Stelle's gravity mentioned before; they are plagued by Ostrogradsky ghosts \cite{Ostrogradsky:1850fid, Pagani:1987ue, Stelle:1977ry, Biswas:2016egy, Ganz:2020skf}. 
The presence of ghosts could break unitarity due to negative norm states \cite{Faddeev:1967fc, Lee:1969fy, Ilhan:2013xe}, although, one can argue that ghosts can be harmless in some cases \cite{Hawking:2001yt, Smilga:2004cy, Smilga:2013vba, Salles:2014rua, Peter:2017xxf, deOSalles:2018eon, Deffayet:2021nnt}. In particular, the unitarity of the standard LM theory has been studied briefly in Ref. \cite{McKeon:2021qhv} through indefinite metric quantization \cite{pauli:1943, Sudarshan:1961vs}. 

In this paper, we propose a modification of the standard LM theory to circumvent these issues in a simpler manner.
To this purpose, we notice that the path integral of the standard LM theory lacks field redefinition invariance, which is an expected property of the path integral of any quantum field theory, including quantum gravity \cite{Arzt:1993gz, Passarino:2016saj, Criado:2018sdb, Casadio:2022ozp}. It is the field redefinition invariance that guarantees, for instance, the validity of the equivalence theorem of the $S$-matrix within the framework of path integral quantization \cite{Chisholm:1961tha, Kamefuchi:1961sb, Salam:1970fso}. Hence, we introduce a determinant factor in the measure of the path integral of the standard LM theory
to restore this invariance. We shall refer to this field redefinition invariant LM formalism as the \emph{modified LM theory}. 

We show that by choosing an appropriate determinant factor to obtain a field redefinition invariant LM theory, the doubling of one-loop contributions is absent. Exponentiating this determinant factor ghost-like fields arise. 
These ghost fields are responsible for the cancellation of the additional one-loop contributions due to the LM fields at the perturbative level. 
However, the main feature of the standard LM theory, namely the restriction of the loop expansion to one-loop order, is not altered by their presence. 
A similar modification of the measure of the path integral is required to retain general covariance in the worldline formalism, which leads to the introduction of the Lee--Yang ghost fields \cite{Bastianelli:1991be, Bastianelli:1998jb} (for reviews, see \cite{Schubert:2001he, Edwards:2019eby}). In this regard, the ghost fields of the modified LM theory are analogous to Lee--Yang ghost fields.

We also argue that the LM fields degrees of freedom cancel against the degrees of freedom of the ghost fields of the modified LM theory, which is similar to the cancellation of the unphysical degrees of freedom of gauge fields by Faddeev--Popov ghosts in gauge theories \cite{Faddeev:1967fc, weinberg:1995i}. To rigorously derive this result, it is necessary to investigate the classical Hamiltonian dynamics of the modified LM theory. This is outside the scope of this paper, which aims to study the quantum LM theory; nevertheless, we briefly discuss the possible nature of the classical modified LM theory at the end of this paper. 

This paper is organized as follows. We review the standard LM theory in Section~\ref{section:SLMT}. In Section~\ref{section:FDofQPI}, we investigate the field redefinition invariance of the LM theory concluding that the path integral of the standard LM theory is not invariant under it. Then, we propose a field redefinition invariant LM theory.  In Section~\ref{section:modLM}, the modified LM theory is studied in more detail. We show that the doubling of the one-loop contribution (and degrees of freedom) is absent, and a diagrammatic analysis is provided. In Section~\ref{section:discussion}, we present a discussion of our proposal and related topics such as unitarity. 
In~\ref{section:FSandGofLM}, we extend the modified LM theory for fermionic systems and suggest a general formulation for systems with both bosonic and fermionic fields. 
Finally, in~\ref{section:symmetries}, we make explicit some of the symmetries that the modified LM theory possesses. 

\section{Standard LM theory}\label{section:SLMT}

In this section, we briefly review the standard LM theory \cite{Brandt:2018lbe, Brandt:2019ymg, Brandt:2020gms}. Consider the action 
\begin{equation}\label{eq:actionPhi}
    S[\phi ] = \int \mathop{d x}  \mathcal{L}  ( \phi ),
\end{equation}
where $ \mathcal{L}  ( \phi  ) $ is the Lagrangian of a set of bosonic fields $ \phi_{i} $ (represented simply by $ \phi $). The path integral quantization procedure yields\footnote{The Lagrangian in Eq.~\eqref{eq:actionPhi} is not singular. It is also assumed that the system is not constrained. }
\begin{equation}\label{eq:GenFunc}
    Z[j] = \int \mathop{\mathcal{D} \phi} \exp \frac{i}{\hbar} \int \mathop{dx} \left( \mathcal{L} ( \phi ) + j \phi \right), \quad (j \phi \equiv j_i \phi_i).
\end{equation}

The standard LM theory is obtained using a Lagrange multiplier field $\lambda$ to restrict the path integral in Eq.~\eqref{eq:GenFunc} to field configurations of $\phi$ that satisfy the classical equations of motion.  
The path integral quantization of the action of Eq. (\ref{eq:actionPhi}) in the framework of the LM theory becomes (sources will not be considered for simplicity\footnote{They can be added straightforwardly. We refer the interested reader to section II of \cite{Brandt:2019ymg}. An interesting example was shown in the section IV of \cite{Brandt:2020gms}.}) \cite{Brandt:2019ymg}
\begin{equation}\label{eq:GenFuncLM}
    Z_{\text{LM}} [0] = \int \mathop{\mathcal D \phi}\mathop{\mathcal D \lambda }\exp \frac{i}{\hbar} \int \mathop{dx} \left( \mathcal{L}(\phi) + \lambda \frac{\delta  S[\phi]}{\delta \phi} 
    \right).
\end{equation}

The functional integral over $\lambda$ in Eq. (\ref{eq:GenFuncLM}) results in the functional $\delta$-function, thus 
\begin{equation}\label{eq:GenFuncDeltaLM_}
Z_{\text{LM}} [0] = 
    \int \mathop{\mathcal{D} \phi} \delta\left ( \frac{\delta S [\phi ]}{\delta \phi} \right )
    \exp \frac{i}{\hbar} S [ \phi ].
\end{equation}
Using the functional analog of 
\begin{equation}\label{A.16}
    \int \mathop{dx} \delta (g(x)) f(x) = \sum_{\bar{x}} |g^\prime (\bar{x})|^{-1} f(\bar{x}), 
\end{equation}
reduces Eq. (\ref{eq:GenFuncDeltaLM_}) to
\begin{equation}\label{A.17}
Z_{\text{LM}} [0]= \sum_{\bar{\phi}(x)} \det\left( \mathcal{L}^{\prime\prime}(\bar{\phi})\right)^{-1}\exp \frac{i}{\hbar}  \int \mathop{dx}  \mathcal{L}(\bar{\phi}).
\end{equation}
In Eq.~\eqref{A.16}, $\bar{x}$ is a solution to 
\begin{equation}\label{A.18}
g(\bar{x}) = 0
\end{equation}
while  $\bar{\phi}(x)$, in Eq.~\eqref{A.17}, satisfies
\begin{equation}\label{A.19}
\left.\frac{\delta S[ \phi ]}{\delta \phi}  \right |_{\phi = \bar{\phi}(x)}  =0.
\end{equation}
The exponential in Eq.~\eqref{A.17} leads to tree diagrams while the determinant yields one-loop contributions, which are twice the contributions obtained with the path integral in Eq.~\eqref{eq:GenFunc}. 

We can see it quantitatively as follows. Comparing the one-loop approximation for the generating functional in Eq.~\eqref{eq:GenFunc}, 
\begin{equation}\label{eq:ApproximationGenFunPhi}
    Z|_{\text{1loop}} = \det\left( \mathcal{L}^{\prime\prime}({\phi})\right)^{-1/2},
\end{equation}
with the exact form obtained in the standard LM theory (in Eq.~\eqref{A.17}) we see that the LM theory results in the square of determinant in Eq.~\eqref{eq:ApproximationGenFunPhi}. Now, using   
the connected generating functional $ W[j] = -i \hbar \ln Z[j] $, we have that (sources are omitted)
\begin{equation}\label{eq:FactorOf2a}
    \begin{split}
        W|_{\text{1loop}} &= - i \hbar \ln Z|_{\text{1loop}}
                      \\ &=  \frac{i \hbar}{2} \mathop{\rm Tr}  \ln {\mathcal{L}}_{\text{}}^{\prime\prime} ( {\phi} ) =  \frac{1}{2}  W_{\text{LM} } |_{\text{1loop}},
    \end{split}
\end{equation}
where $ W_{\text{LM}} = -i \hbar \ln Z_{\text{LM}} [0] $. Hence, 
\begin{equation}\label{eq:FactorOf2}
     W_{\text{LM}}|_{\text{1loop}}  = 2 W|_{\text{1loop}},
\end{equation}
which shows that in the standard LM theory the one-loop contributions are doubled. 
The factor of $2$ in Eq.~\eqref{eq:FactorOf2} comes from extra contributions due to the LM field $ \lambda $ in Eq.~\eqref{eq:GenFuncLM}. 


Following the procedure of Refs. \cite{Brandt:2019ymg, Brandt:2020gms}, we provide a diagrammatic analysis. Assuming that the Lagrangian 
\eqref{eq:GenFuncLM} has the polynomial form (field indices are explicit)
\begin{equation}\label{eq:expandedAction}
    \mathcal{L}  ( \phi ) = 
    \frac{1}{2!} a_{i j}^{(2)} \phi_{i} \phi_{j} + \frac{1}{3!} a_{ijk}^{(3)} \phi_{i} \phi_{j} \phi_{k} +
    \frac{1}{4!} a_{ijkl}^{(4)} \phi_{i} {\phi}_{j}^{} \phi_{k} \phi_{l} +   \cdots,
\end{equation}
 we have that 
\begin{equation}\label{eq:expandedLM_action}
    \begin{split}
        \mathcal{L} + \lambda \frac{\partial \mathcal{L}}{\partial \phi} =  
        \frac{1}{2!} & a_{i j}^{(2)} \phi_{i} \phi_{j} + \frac{1}{3!} a_{ijk}^{(3)} \phi_{i} \phi_{j} \phi_{k}   +   
        \frac{1}{4!} a_{ijkl}^{(4)} \phi_{i} {\phi}_{j}^{} \phi_{k} \phi_{l} +   \cdots \\ &  +  a_{i j}^{(2)} \phi_{i} \lambda_{j} + 
                       \frac{1}{2!} a_{ijk}^{(3)} \phi_{i} \phi_{j} \lambda_{k} +  
    \frac{1}{3!} a_{ijkl}^{(4)} \phi_{i} {\phi}_{j}^{} \phi_{k} \lambda_{l} +   \cdots,
    \end{split}
\end{equation}
and the Feynman rules can be obtained straightforwardly. As we can see in Eq.~\eqref{eq:expandedLM_action}, along with the vertices that come from the original Lagrangian $ \mathcal{L} ( \phi ) $ we have also similar vertices in which one field $ \phi $ is replaced by an LM field $ \lambda $. We obtain the propagators inverting the matrix of the bilinear terms in $ \phi $ and $ \lambda $:
\begin{equation}\label{eq:def:matrixProp}
    \begin{pmatrix}
        a_{ij} & a_{ij} \\
        a_{ij} & 0
        \end{pmatrix}^{-1} = 
     \begin{pmatrix}
         0 & a_{ij}^{-1}  \\
         a_{ij}^{-1}& - a_{ij}^{-1} 
        \end{pmatrix},
\end{equation}
which reveals that there is no propagator $ \left \langle \phi_{i} \phi_{j} \right\rangle  $ for the field $ \phi $. Instead there are the mixed propagators $ \langle \phi_{i} \lambda_{j} \rangle = \left \langle \lambda_{i} \phi_{j} \right\rangle =-i a_{ij}^{-1} $, and the propagator $ \left \langle \lambda_{i} \lambda_{j}\right\rangle = i a_{ij}^{-1}$ of the field $ \lambda $ that gets a negative sign.

\begin{figure*}[ht]
    \centering
    \includegraphics[scale=0.8]{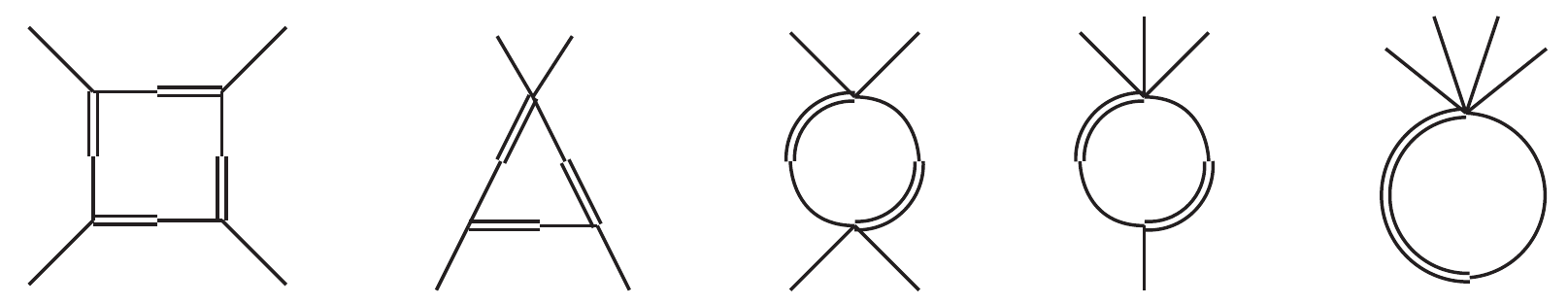}
    \caption{One-loop contributions to 4-point amplitude $ ( \phi_{i} \phi_{j} \phi_{k} \phi_{l} )$ in the LM theory. Solid and doubled lines represent respectively the fields $ \phi $ and $ \lambda $.}
    \label{fig:1}
\end{figure*}

\begin{figure*}[ht]
    \centering
    \includegraphics[scale=0.60]{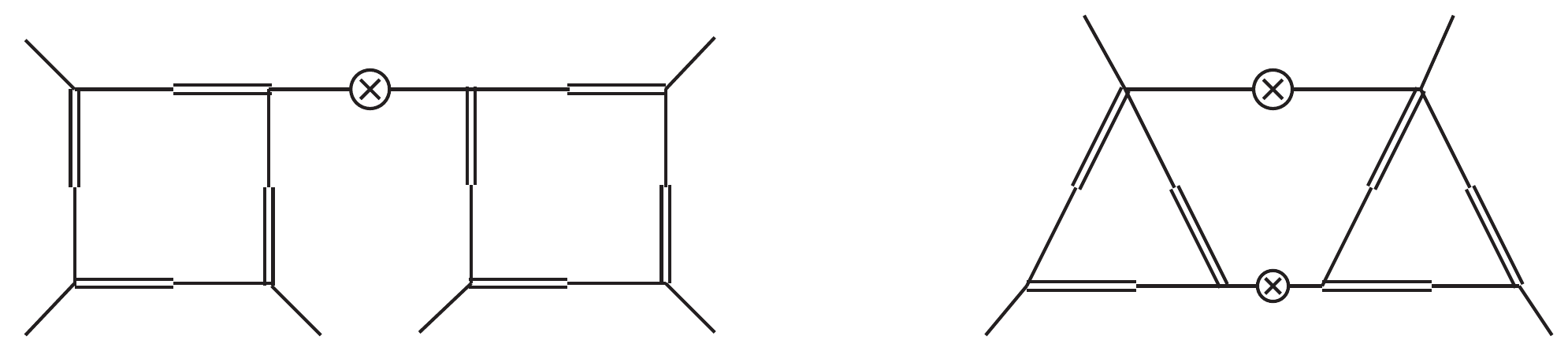}
    \caption{Diagrams higher than one loop order cannot be drawn joining one-loop diagrams in the LM theory. In these examples crossed lines represent forbidden topologies.}
    \label{fig:2a}
\end{figure*}

\begin{figure*}[ht]
    \centering
    \includegraphics[scale=0.3]{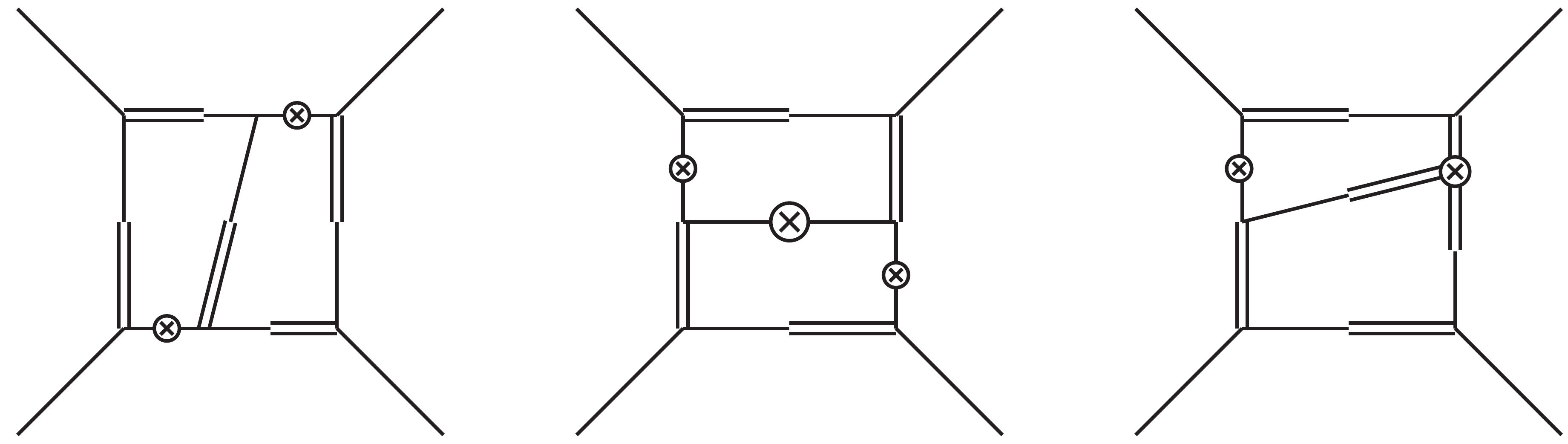}
    \caption{Higher-loop order diagrams cannot be drawn at all in the LM theory. Crosses denote forbidden topologies.}
    \label{fig:2}
\end{figure*}

By these rules, we cannot draw any diagram beyond one-loop order. We can derive it as follows. 
\begin{enumerate}[(i)]
    \item To draw a one-loop diagram with $ \lambda $ in an external leg requires at least one internal field $ \phi $ line since there is no vertex with more than one LM field $ \lambda $. There is no propagator for the field $ \phi $, therefore it is not possible to draw these diagrams. 

    \item We can only draw one-loop diagrams with $ \phi $ in the external lines, as we see in Fig.~\ref{fig:1}, and only mixed propagators can appear in the internal lines. 

    \item  These one-loop diagrams cannot be iterated to construct any diagram of higher-loop order (see Fig.~\ref{fig:2a}), since there is no propagator for the field $ \phi $. Other diagrams of higher order also cannot be drawn for similar, see Fig.~\ref{fig:2}.
\end{enumerate}


\begin{figure*}[ht]
    \centering
    \includegraphics[scale=0.8]{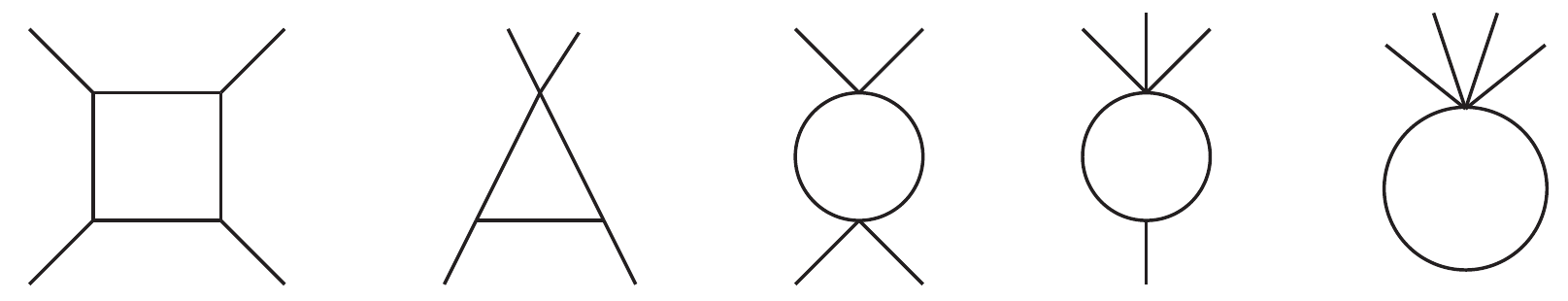}
\caption{One-loop contributions to 4-point amplitude $ ( \phi_{i} \phi_{j} \phi_{k} \phi_{l} )$ in the theory described by the Lagrangian $ \mathcal{L} ( \phi )$ in Eq.~\eqref{eq:expandedAction}.}
    \label{fig:3}
\end{figure*}

\begin{figure}[ht]
    \centering
    \includegraphics[scale=0.55]{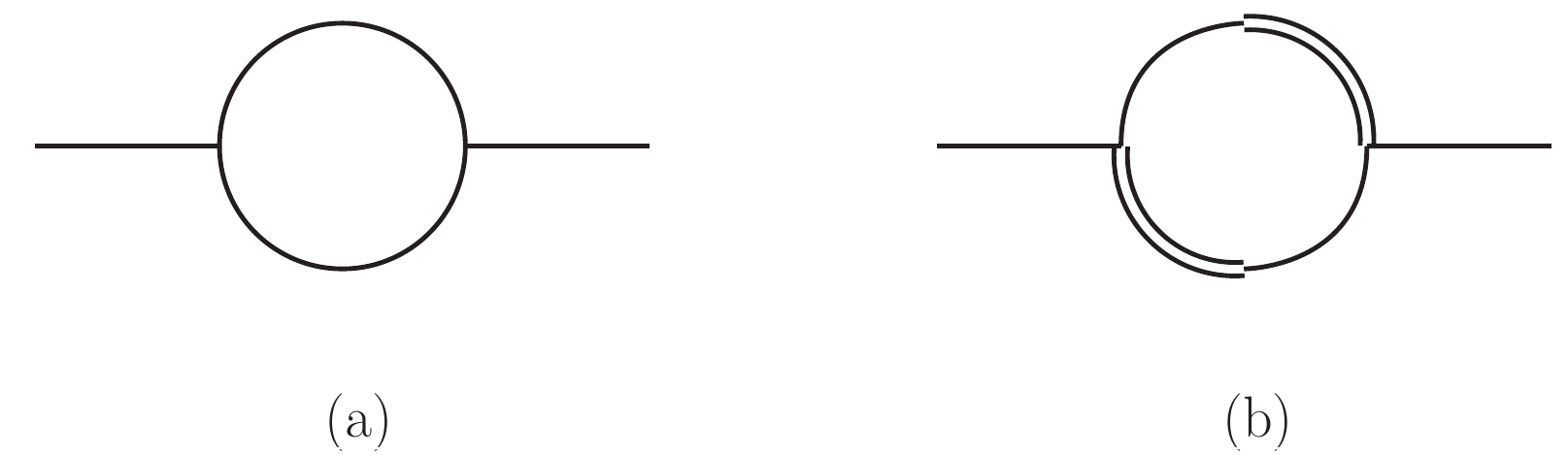}
    \caption{Diagrams (a) and (b) are respectively the one-loop contributions to 2-point amplitude $( \phi_{i} \phi_{j} )$ in the theory described by the Lagrangians in the Eqs.~\eqref{eq:expandedAction} and \eqref{eq:expandedLM_action} (LM theory).}
    \label{fig:4}
\end{figure}

Comparing the diagrams of the LM theory, see Fig.~\ref{fig:1}, to the diagrams in Fig.~\ref{fig:3} from the original theory described by the Lagrangian in Eq.~\eqref{eq:expandedAction}, we see that the one-loop contributions are doubled in the LM theory described by the Lagrangian in Eq.~\eqref{eq:expandedLM_action}. 
This doubling appears in the diagrams as a relative combinatorial factor of 2, which is consistent with Eq.~\eqref{eq:FactorOf2}. For instance, while the symmetry factor of the diagram in Fig.~\ref{fig:4}(a) is $2$, the diagram in Fig.~\ref{fig:4}(b) has a symmetry factor of $1$. 
Thus, the LM theory contribution shown in Fig.~\ref{fig:4}(b) results in twice the usual one-loop contribution shown in Fig.~\ref{fig:4}(a). 

\section{Field redefinitions in quantum path integrals} \label{section:FDofQPI}

We start this section by reviewing the invariance of quantum path integral under field redefinitions, which is an expected property of any quantum field theory described within the path integral quantization procedure. We will present it in a general form in terms of the generating functional \eqref{eq:GenFunc}, obtained in the path integral quantization of a general theory described by the action in Eq.~\eqref{eq:actionPhi}. 

One can redefine the field $ \phi $ so that the generating functional in Eq.~\eqref{eq:GenFunc} remains invariant. Under the local redefinition of the fields 
\begin{equation}\label{eq:RedFieldPhi}
    \phi \rightarrow \phi' =  F[ \phi ],
\end{equation}
where $ F [ \phi ] $ is an invertible functional, the generating functional in Eq.~\eqref{eq:GenFunc} reads 
\begin{equation}\label{eq:GenFuncPHI}
    Z[j] = \int \mathop{\mathcal{D} \phi'} \det \left ( \frac{\delta \phi  }{\delta \phi'}\right ) \exp \frac{i}{\hbar} \int \mathop{dx} \left( \mathcal{L} ( F^{-1}[ \phi'] ) + j F^{-1} [ \phi'] \right),
\end{equation}
where
\begin{equation}\label{eq:JacobianF}
     \det \left ( \frac{\delta F^{-1} [ \phi '] }{\delta \phi '}\right )=  \det  \frac{\delta \phi}{\delta \phi'} 
    \equiv \det\bm{J}_{\phi}
\end{equation}
is the Jacobian determinant of the redefinition \eqref{eq:RedFieldPhi}. This Jacobian determinant can be sufficiently general. The only assumptions are the non-singular condition $ \det \bm{J}_{\phi} \neq 0 $ and the absence of non-local terms which could be generated by the field redefinition given by Eq.~\eqref{eq:RedFieldPhi}.

Hence, the quantum path integral is invariant under field redefinitions as long its measure is supplemented by the corresponding Jacobian determinant, and the source term properly altered \cite{Criado:2018sdb}. For non-linear field redefinitions, it may also be necessary to introduce an extra term in the action \cite{Apfeldorf:2000dn}, which is not relevant to this paper. For a fermionic field (Grassmann field), we would get the inverse of the Jacobian factor, for more detail, see~\ref{section:FSandGofLM}.

Note that in this paper we only consider quantum path integrals. Then, from now on, we use ``path integral'' to refer to quantum path integral.
The classical path integral, firstly introduced in \cite{Gozzi:1986ge, Gozzi:1989bf} as the functional formulation\footnote{For a review, see Ref. \cite{Gozzi:2001nm}.} of the Koopman, von Neumann theory \cite{koopman:1931, v.neumann:1932} (the operational version of classic mechanics), does not behave in the same manner under field redefinitions. For the classical path integral, the Jacobian determinant is absent. 

In the next section, we will see that the behavior of the path integral of the standard LM theory under field redefinitions differs from both, classical and quantum. Hence, to restore the quantum behavior we propose to introduce a new term in the measure of the path integral of the LM theory.

\subsection{Field redefinitions in the standard Lagrange multiplier theory}

Let us use the generating functional in Eq.~\eqref{eq:GenFuncDeltaLM_}  so that we can see transparently how the generating functional of the standard LM theory transforms under field redefinitions. We also provide an alternative derivation using the generating functional in its standard form, that is Eq.~\eqref{eq:GenFuncLM}.

Under the field redefinition \eqref{eq:RedFieldPhi}, the measure of integration of the generating functional of the standard LM theory \eqref{eq:GenFuncDeltaLM_} transforms as 
\begin{equation}\label{eq:MeasureIntLM_}
    \mathop{\mathcal{D} \phi} \delta \left( \frac{\delta S[\phi ] }{\delta \phi} \right) \rightarrow 
    \mathop{\mathcal{D} \phi'}
    \det \left(\frac{\delta \phi  }{\delta \phi'}  \right)\delta  \left ( \frac{\delta \phi ' }{\delta \phi } 
    \frac{\delta S'[\phi']}{\delta \phi'} \right ) = \mathcal{D} \phi '  \det \bm{J}_{\phi}^2 \delta \left ( \frac{\delta S'[\phi']}{\delta \phi '} \right ), 
\end{equation}
where we have used the following property of the delta function:
\begin{equation}\label{eq:Delta}
    \begin{split}
        \delta\left ( \frac{\delta \phi ' }{\delta \phi } \frac{\delta S'[\phi'] }{\delta \phi'} \right) &= 
        {\left\|\frac{\delta \phi ' }{\delta \phi}\right\|^{-1}} 
    {\delta \left ( \frac{\delta S'[\phi'] }{\delta \phi '}\right )}
                                                                                                        \\ &=  \det \bm{J}_{\phi} \delta \left ( \frac{\delta S'[\phi'] }{\delta \phi '} \right ) 
\end{split}
\end{equation}
(since the transformation in Eq.~\eqref{eq:RedFieldPhi} is invertible).
Note that from Eq.~\eqref{eq:GenFuncPHI} one infers that the transformed action $ S' [\phi'] = S [ F^{-1}[ \phi ']]$. 

Therefore, the transformation of the generating functional of the standard LM theory reads 
\begin{equation}\label{eq:GenFuncLM_PHI2_2}
    Z_{\text{LM}}[0] \rightarrow 
    Z_{\text{LM}} '[0] = \int \mathop{\mathcal{D} \phi'} \det \bm{J}_{\phi}^2 
\delta\left(\frac{\delta S'[\phi']}{\delta \phi'} \right)
    \exp \frac{i}{\hbar} S'[\phi'].
\end{equation}
This transformation deviates from the expected behavior of path integrals under field redefinitions, which we presented in general form in Eq.~\eqref{eq:GenFuncPHI}. In the framework of the standard LM theory the Eq.~\eqref{eq:GenFuncPHI}  is given by
\begin{equation}\label{eq:GenFuncLM_PHI2_3}
 \int \mathop{\mathcal{D} \phi'} 
 \det \bm{J}_{\phi} \delta \left(\frac{\delta S'[\phi']}{\delta \phi'} \right)
    \exp \frac{i}{\hbar} S'[\phi'],
\end{equation}
which differs from Eq.~\eqref{eq:GenFuncLM_PHI2_2} by having the extra Jacobian factor $ \det \bm{J}_{\phi}$. Therefore, we can conclude that the standard LM theory is not invariant under field redefinitions. 

One can argue that the extra factor in Eq.~\eqref{eq:GenFuncLM_PHI2_2} is expected since in the LM theory any redefinition of the field $ \phi $ must be accompanied by a redefinition of the respective LM field $ \lambda $. 
This is necessary to preserve the form of the LM theory. 
Indeed, the field redefinition \eqref{eq:RedFieldPhi} in the generating functional \eqref{eq:GenFuncLM} must be supplemented by an appropriate redefinition of the LM field $ \lambda $: 
\begin{equation}\label{eq:FRofL}
    \lambda \to \lambda ' = \lambda \frac{\delta \phi '}{\delta \phi},
\end{equation}
so that
\begin{equation}\label{eq:StructureLM}
    S[ \phi ] + \lambda \frac{\delta S[\phi ]}{\delta \phi} \to S'[\phi'] + \lambda ' \frac{\delta S'[\phi']}{\delta \phi '}.
\end{equation}
(The special case of gauge invariance is treated in section V of Ref.~\cite{Brandt:2020gms}.)
It is this companion redefinition that contributes to the extra Jacobian factor in Eq.~\eqref{eq:GenFuncLM_PHI2_2} since 
\begin{equation}\label{eq:TofLMdiff}
    \mathop{\mathcal{D} \lambda} \to \mathop{\mathcal{D} \lambda'} \det \frac{\delta \lambda}{\delta \lambda '}  = \mathop{\mathcal{D} \lambda '} \det \bm{J}_{\phi}.
\end{equation}
This is another approach to derive the result in Eq.~\eqref{eq:MeasureIntLM_}.

Although the extra factor in Eq.~\eqref{eq:GenFuncLM_PHI2_2} is necessary to the form invariance of the action of the standard LM theory, the behavior of the measure of its path integral under field redefinitions diverges from the expected (see Eq.~\eqref{eq:GenFuncLM_PHI2_3}), which indicates an inconsistency since the extra  Jacobian factor does not appear under redefinitions of the LM field $ \lambda $. If we redefine, independently, the field $ \phi $ and its LM field $ {\lambda} $
leading to the respective Jacobian determinants $ \bm{J}_{\phi } $, $ \bm{K}_{\lambda} $; then the measure of integration in Eq.~\eqref{eq:GenFuncLM} transform as 
\begin{equation}\label{eq:TofMI_LM}
    \mathop{\mathcal{D} \phi} 
    \mathop{\mathcal{D} \lambda} 
    \rightarrow 
\mathop{\mathcal{D} \phi'}
    \mathop{\mathcal{D} \lambda '} 
    \det \bm{J}_{\phi} \bm{J}_{\lambda} 
    \rightarrow 
\mathop{\mathcal{D} \phi'}
    \mathop{\mathcal{D} \lambda ''} 
    \det \bm{J}_{\phi}^{2} \bm{K}_{\lambda'}
\end{equation}
in which we have used the Eq.~\eqref{eq:TofLMdiff}.
This shows the discrepancy between the factors of the Jacobian determinant for $ \phi $ and its LM field $ \lambda $.

One could argue that imposing $ \bm{K}_{\lambda'} = \bm{J}_{ \phi }^{-1} $ would lead to a field redefinition invariant path integral for the standard LM theory, since $ \det \bm{J}_{\phi}^{2} \bm{K}_{\lambda'} = \det \bm{J}_{\phi} $. However, these field redefinitions of the LM field $ \lambda $ would break the form of the action of the LM theory shown in Eq.~\eqref{eq:StructureLM}. Note that it does not occur for the companion redefinition of the LM field in Eq.~\eqref{eq:FRofL}; the linearity of the LM field $ \lambda $ is kept unaltered (see Eq.~\eqref{eq:FRofL}).

\subsection{Field redefinition invariant LM theory}

In the previous section, we concluded that the standard LM theory is not invariant under field redefinitions. In this section, we show that the invariance under field redefinitions can be restored if we modify the measure of integration of the standard LM theory. 

Let us consider the following measure of integration for the LM theory, where we introduce a term $\Delta[ \phi ]$, 
\begin{equation}\label{eq:ModifiedLM_Delta}
\mathop{\mathcal{D} \phi} \Delta[ \phi ] \delta \left( \frac{\delta S [\phi ] }{\delta \phi} \right)
\end{equation}
(compare with the usual in Eq.~\eqref{eq:GenFuncDeltaLM_}). Under the field redefinitions in Eq.~\eqref{eq:RedFieldPhi} the modified measure transforms as 
\begin{equation}\label{eq:ModifiedLM_DeltaTransforms}
\mathop{\mathcal{D} \phi} \Delta[ \phi ] \delta \left( \frac{\delta S [\phi ] }{\delta \phi} \right) \rightarrow
\mathop{\mathcal{D} \phi'} \det \bm{J}_{\phi}^2 \Delta'[ \phi'] \delta \left( \frac{\delta S'[\phi'] }{\delta \phi'} \right). 
\end{equation}
Thus, provided that 
\begin{equation}\label{eq:DELTAtransform}
    \Delta'[\phi']= \det \bm{J}_{\phi}^{-1} \Delta[\phi']
\end{equation}
the measure of the modified LM formalism transforms appropriately (as shown in Eqs.~\eqref{eq:GenFuncPHI} and \eqref{eq:GenFuncLM_PHI2_3}) as 
\begin{equation}\label{eq:ModifiedLM_DeltaTransforms2}
\mathop{\mathcal{D} \phi} \Delta[ \phi ] \delta \left( \frac{\delta S [\phi ] }{\delta \phi} \right) \rightarrow
\mathop{\mathcal{D} \phi'} \det \bm{J}_{\phi} \Delta[ \phi'] \delta \left( \frac{\delta S'[\phi'] }{\delta \phi'} \right)
\end{equation}
under the field redefinition \eqref{eq:RedFieldPhi}.
Consequently, the path integral of the LM theory with the modified measure \eqref{eq:ModifiedLM_Delta} shall now be invariant under field redefinitions.

Our choice for $ \Delta [\phi ]$ reads 
\begin{equation}\label{eq:def:Delta}
        \Delta [ \phi ] 
        = \det \left ( \frac{\delta^{2} S[ \phi ] }{\delta \phi \delta \phi}\right )^{+1/2},
\end{equation}
that is, the Pfaffian of the Hessian of the action \eqref{eq:actionPhi}. Note that Eq.~\eqref{eq:def:Delta} is the inverse of the one-loop quantum corrections, which is shown in Eq.~\eqref{eq:ApproximationGenFunPhi}. Not only this determinant will lead to non-trivial contributions to the perturbative expansion of the LM theory that are responsible for canceling unphysical contributions due to the LM field $ \lambda $, but will also restore the proper behavior of its path integral under field redefinitions. 

It is worth mentioning that in Eq.~\eqref{eq:def:Delta} it should be the absolute value of the Hessian, but we assume that the Hessian of the action \eqref{eq:actionPhi} is positive\footnote{The determinant \eqref{eq:def:Delta} in the Hamiltonian formulation is known to be always positive \cite{Gozzi:1989bf, Gozzi:2001nm}. This assumption also lets us avoid some subtleties that can appear due to phases \cite{McKeon:1992np}.} so that the absolute value can be omitted in Eq.~\eqref{eq:def:Delta}. In ~\ref{section:FSandGofLM} we extend our proposal to fermionic systems and suggest a generalization of the Eq.~\eqref{eq:def:Delta}.

Now, let us show that $ \Delta [ \phi ] $ in Eq.~\eqref{eq:def:Delta} is consistent with the transformation law in Eq.~\eqref{eq:DELTAtransform}. 
The functional derivative transforms as 
\begin{equation}\label{eq:transDerivative}
    \frac{\delta }{\delta \phi} = \frac{\delta \phi ' }{\delta \phi} \frac{\delta }{\delta \phi '} = \bm{J}_{\phi}^{-1} \frac{\delta }{\delta \phi '}  
\end{equation}
thus
\begin{equation}\label{eq:trans2Derivative}
    \frac{\delta^{2} }{\delta \phi \delta \phi} 
    =\bm{J}_{\phi}^{-2} \left(\frac{\delta^{2} }{\delta \phi ' \delta \phi '} -   \frac{\delta \bm{J}_{\phi} }{\delta \phi' } \bm{J}_{\phi}^{-1}\frac{\delta}{\delta \phi '}\right).
\end{equation}
Therefore,
\begin{equation}\label{eq:transDelta}
    \Delta'[\phi ']
    = \det \bm{J}_{\phi}^{-1} \Delta [ \phi '] \det K^{+1/2} 
\end{equation}
with
\begin{equation}\label{eq:def:K}
    K = 1 - \left(\frac{\delta^{2} S'[\phi']}{\delta \phi ' \delta \phi '}\right)^{-1}  \frac{\delta \bm{J}_{\phi}}{\delta \phi ' } \bm{J}_{\phi}^{-1}\frac{\delta S'[\phi']}{\delta \phi'}.
\end{equation}
The factor $ \det K^{+1/2} $ in Eq.~\eqref{eq:transDelta} implies a violation of the transformation law \eqref{eq:DELTAtransform}, since, in general, $ \delta \bm{J}_{\phi} / \delta \phi'$ does not vanish. However, it does not contribute to the resulting measure of integration of the LM theory in Eq.~\eqref{eq:ModifiedLM_DeltaTransforms}. Substituting Eq.~\eqref{eq:transDelta} in the left hand side of Eq.~\eqref{eq:ModifiedLM_DeltaTransforms}: 
\begin{equation}\label{eq:transDelta1}
    \begin{split}
        \mathop{\mathcal{D} \phi'} \det \bm{J}_{\phi}^{2} \Delta'[\phi '] \mathop{\delta} \left ( \frac{\delta S'[\phi'] }{\delta \phi'}\right )  
    ={}&  \mathop{\mathcal{D} \phi'}\det \bm{J}_{\phi} \Delta[\phi ']
   \det K^{+1/2}  \mathop{\delta} \left ( \frac{\delta S'[\phi'] }{\delta \phi'} \right ) \\ 
    ={}& \mathop{\mathcal{D} \phi'} \det \bm{J}_{\phi} \Delta[\phi '] \mathop{\delta} \left ( \frac{\delta S'[\phi'] }{\delta \phi'}\right )
\end{split}
\end{equation}
yields the left hand side of Eq.~\eqref{eq:ModifiedLM_DeltaTransforms2}, since we have that $\det K = \det 1$ by the delta function. 
Therefore, the proposed definition of $ \Delta [ \phi ]$ shown in Eq.~\eqref{eq:def:Delta} satisfies the transformation law \eqref{eq:DELTAtransform} in the LM formalism.

Other possibilities that satisfy the transformation law \eqref{eq:DELTAtransform} could be considered. However, our proposal \eqref{eq:def:Delta} is the only one that leads to a field redefinition invariant formalism that produces the expected tree and one-loop effects.

Let us consider some scenarios. For instance, the measure   
\begin{equation}\label{eq:ModifiedLM_Deltap}
    \mathop{\mathcal{D} \phi} \Delta[ \phi ]^{p} \delta \left( \frac{\delta S [\phi ] }{\delta \phi} \right),
\end{equation}
where $p$ is an integer and $ \Delta [ \phi ] $ is given by Eq.~\eqref{eq:def:Delta}.
\begin{enumerate}[(i)]
    \item  When $ p=0$, we have the standard LM formalism: the one-loop contributions are doubled, and its path integral is not field redefinition invariant. 
    \item  If $ p=1$, we obtain the proposed field redefinition invariant formalism in which one-loop contributions are correctly produced and tree-level effects are kept unaltered. 
    \item With $ p =2 $, all quantum effects are eliminated. The path integral will be similar to the classical path integral proposed by Gozzi \cite{Gozzi:1986ge}. This path integral is not field redefinition invariant and the Jacobian factor expected (cf. Eq.~\eqref{eq:GenFuncLM_PHI2_3}) is now absent.
\end{enumerate}
The only value of $ p$ that satisfies the transformation law shown in Eq.~\eqref{eq:DELTAtransform} is $ p=1$, which is the one proposed here (ii). 

Although the transformation law \eqref{eq:DELTAtransform} still left us with some arbitrariness to define $ \Delta [ \phi ] $, by using \eqref{eq:def:Delta} we get to remove unphysical contributions due to the LM field $ \lambda $. These unphysical contributions (and the negative norm states) found in the standard LM theory can violate unitarity. However, this is avoided in the modified LM theory by defining $ \Delta [ \phi ] $ as \eqref{eq:def:Delta}. 
(In the standard LM theory, it can be resolved using indefinite metric quantization \cite{McKeon:2021qhv}.) 
Thus, the proposed definition in Eq.~\eqref{eq:def:Delta} is the only choice that is consistent with the transformation law \eqref{eq:DELTAtransform} and unitarity.

Using Eq.~\eqref{eq:def:Delta}, the path integral of the modified LM theory reads 
\begin{equation}\label{eq:GenFuncLM_Modified}
    \mathcal{Z}_{\text{LM}}[0] = \int \mathop{\mathcal{D} \phi} \mathop{\mathcal{D} \lambda} \det \left ( \frac{\delta^{2} S [\phi]}{\delta \phi \delta \phi}\right )^{+1/2} \exp \frac{i}{\hbar}  S_{\text{LM}} [\phi ],
\end{equation}
where 
\begin{equation}\label{eq:def:S_LM}
    S_{ \text{LM} } [\phi ]= 
    \int \mathop{dx}\left( \mathcal{L} ( \phi ) + \lambda  \frac{\delta S [\phi]   }{\delta \phi} \right)
\end{equation}
is the action of the standard LM theory. The path integral in Eq.~\eqref{eq:GenFuncLM_Modified} (invariant under field redefinitions) is our proposal for the quantum LM theory.

We can write the determinant in Eq.~\eqref{eq:GenFuncLM_Modified} locally with the introduction of new fields. For this, we first rewrite it as
\begin{equation}\label{eq:DeltaInTermsOfFieldsBefore}
    \begin{split}
        \det \left ( \frac{\delta^{2} S [\phi] }{\delta \phi \delta \phi}\right )^{+1/2} =
                     \det \left(\frac{\delta^{2} S_{} [ \phi ] }{\delta \phi \delta \phi}\right)\det \left(\frac{\delta^{2} S [ \phi ] }{\delta \phi \delta \phi}\right)^{-1/2}   
    \end{split} 
\end{equation}
and then exponentiate the determinants in Eq.~\eqref{eq:DeltaInTermsOfFieldsBefore}, which results in 
\begin{equation}\label{eq:DeltaInTermsOfFields}
        \Delta [ \phi ] 
                     =  \int \mathop{\mathcal{D}{\bar{\theta}}} \mathop{\mathcal{D }\theta } \mathop{\mathcal{D} \chi} \exp \frac{i}{\hbar} \int \mathop{d^{}x} \left ( \bar{\theta} \frac{\delta^{2 }S [\phi ] }{\delta \phi \delta \phi} \theta + \frac{1}{2} \chi \frac{\delta^{2} S [\phi] }{\delta \phi \delta \phi} \chi\right ),
\end{equation}
where $ \bar{\theta} , \theta $ and $\chi$ are scalar fermionic and bosonic fields, respectively. 
An analogous procedure is done in the worldline formalism to obtain a covariant path integral. The term $ \Delta[g] \thicksim \sqrt{g}  $, analogous to Eq.~\eqref{eq:def:Delta}, is required to the measure of the path integral retain general covariance so that the Lee--Yang ghost fields are introduced \cite{Bastianelli:1991be, Bastianelli:1998jb}.

These new fields (analogous to Lee--Yang ghost fields) are ghost-like. The LM field $ \lambda $ leads to negative norm states \cite{McKeon:1992rq, McKeon:2021qhv} as well as the ghost fields $ \theta $, $ \bar{\theta} $ that violate the spin-statistic theorem. The ghost fields $ {\theta} $, $ {\bar{\theta}} $ are similar to Faddeev--Popov ghosts \cite{Faddeev:1967fc}, while the field $ \chi $ plays the role of a third ghost\footnote{A third ghost (the Nielsen-Kallosh ghost \cite{Nielsen:1978mp, Kallosh:1978de}) appears in gauge theories for non-singular gauges with functional dependence \cite{Batalin:1983ar}. It is similar to our case given that the ghost-like fields in the action Eq.~\eqref{eq:DeltaInTermsOfFields} depend on a functional, the Hessian of the action $S[ \phi ]$. Gauge theories  and gravity with high derivative gauges are examples of models that need a third ghost \cite{Brandt:2007td, Brandt:2011zb, Ohta:2020bgz}.}. Although in principle, these two kinds of ghosts (Faddeev--Popov and the Lee--Yang-like $ \theta$, $ \bar{\theta} $, and $ \chi $ ghost fields) are introduced for different reasons, it can be said that both are necessary for an appropriate path integral quantization. 

From Eq.~\eqref{eq:DeltaInTermsOfFields} we have the ghost fields action 
\begin{equation}\label{eq:def:Sgh}
    S_{\text{gh}}[ \phi ] = 
 \int \mathop{dx}\left(  \bar{\theta} \frac{\delta^{2 }S [\phi ] }{\delta \phi \delta \phi} \theta + \frac{1}{2} \chi \frac{\delta^{2} S [\phi] }{\delta \phi \delta \phi} \chi\right)
\end{equation}
that allows us to define an effective action for the modified LM theory as 
\begin{equation}\label{eq:def:Seff}
    \begin{split}
        S_{\text{eff}} [\phi ] &=   S_{\text{LM}}[ \phi ]+ S_{\text{gh}}[ \phi ] \\   
                                                                                              &= \int \mathop{dx}\left( \mathcal{L} ( \phi ) + \lambda  \frac{\delta S
    [\phi ]   }{\delta \phi}  + \bar{\theta} \frac{\delta^{2 }S [\phi ] }{\delta \phi \delta \phi} \theta + \frac{1}{2} \chi \frac{\delta^{2} S [\phi] }{\delta \phi \delta \phi} \chi\right).
\end{split}
\end{equation}
Substituting Eqs.~\eqref{eq:DeltaInTermsOfFieldsBefore}---\eqref{eq:def:Seff} into Eq.~\eqref{eq:GenFuncLM_Modified} yields 
\begin{equation}\label{eq:GenFuncLM_ModifiedLocal}
    \mathcal{Z}_{\text{LM}}[0] = \int \mathop{\mathcal{D} \phi} \mathop{\mathcal{D} \lambda} \mathop{\mathcal{D}\bar{\theta}}  \mathop{\mathcal{D}\theta } \mathop{\mathcal{D} \chi  } 
    \exp \frac{i}{\hbar} S_{\text{eff} } [\phi ],
\end{equation}
which is the proper path integral for the LM theory. 
This means that under the field redefinition \eqref{eq:RedFieldPhi} the generating functional \eqref{eq:GenFuncLM_ModifiedLocal} transform appropriately, 
\begin{equation}\label{eq:transGenFuncmLM_local}
    \mathcal{Z}_{\text{LM}}[0] 
    \to
    \mathcal{Z'}_{\text{LM}}[0] 
    = \int \mathop{\mathcal{D} \phi'} \mathop{\mathcal{D} \lambda'} \mathop{\mathcal{D}\bar{\theta}'}  \mathop{\mathcal{D}\theta'} \mathop{\mathcal{D} \chi'} \det \bm{J}_{\phi}  
    \exp \frac{i}{\hbar} S_{\text{eff} }^\prime [\phi'],
\end{equation}
agreeing with the expected behavior shown in Eqs.~\eqref{eq:GenFuncPHI} and \eqref{eq:GenFuncLM_PHI2_3}. 

In Eq.~\eqref{eq:transGenFuncmLM_local} the bosonic fields $B= ( \phi , \, \lambda , \, \chi )$ lead to Jacobian factors of $ \det \bm{J}_{\phi}$, while fermionic fields $F = ( \theta \, , \bar{\theta} )$ contribute inversely (see~\ref{section:FSandGofLM}) resulting in the expected Jacobian factor. That is, $ \mathop{\mathcal{D} B} = \det \bm{J}_{\phi} \mathop{\mathcal{D} B'}$ and $\mathop{\mathcal{D} F} = \det^{-1} \bm{J}_{\phi} \mathop{\mathcal{D} F'}$ since when the field $ \phi $ undergoes a field redefinition the associated fields $ \lambda $, $ \chi $, $ \theta $, and $ \bar{\theta} $; introduced in the modified LM formalism, must transform accordingly. Thus, a field redefinition of $ \phi $ such as \eqref{eq:RedFieldPhi} must be accompanied by the field redefinitions of the associated fields
\begin{subequations}\label{eq:transformationsofmLM}
    \allowdisplaybreaks
    \begin{align}
        \lambda' ={}& \lambda \frac{\delta F[\phi]}{\delta \phi}, \\
        \chi' ={}& \chi \frac{\delta F[\phi]}{\delta \phi} , \\
        \theta ' ={}& \theta \frac{\delta F[\phi]}{\delta \phi} , \\  \bar{\theta} ' ={}& \bar{\theta} \frac{\delta F[\phi]}{\delta \phi}; 
    \end{align}
\end{subequations}
while other independent fields are kept unaltered, in order to preserve form invariance of the effective action of the modified LM in Eq.~\eqref{eq:def:Seff}. Eq.~\eqref{eq:transformationsofmLM} is the analog of Eq.~\eqref{eq:TofLMdiff} to the modified LM theory. Besides that, for gauge theories, the prescription given by Eq.~\eqref{eq:transformationsofmLM} can be used to extend gauge invariances of the original theory to the modified LM formalism.

The action of the modified LM theory in Eq.~\eqref{eq:def:Seff} is invariant under the ghost number ($ \mathop{gh} $) symmetry \footnote{This symmetry is related to the ghost charge, which is conserved by the Noether theorem. In~\ref{section:symmetries} we also show that the ghost sector is invariant under (anti)BRST-like symmetry due to its natural supersymmetric structure.} presented in~\ref{section:symmetries}.
It implies that the ghost number of the effective action should vanish
$ \mathop{gh} S_{\text{eff}} [\phi ] = 0$, while $ \mathop{gh} [ \theta ]= -\mathop{gh} [ \bar{\theta} ]=1$ and $ \mathop{gh} [\chi ]=0$, which agree with ref.~\cite{Batalin:1983ar}. This shows another similarity between the ghost fields $ \theta $, $ \bar{\theta} $, and Faddeev--Popov ghosts.

By this similarity, let us suppose that the ghost fields $ \theta$ and $ \bar{\theta} $ serve as negative degrees of freedom in the modified LM theory. If we denote the degrees of freedom of a field $ I$ by $ N_I$ and the total degrees of freedom of the fermionic ghost fields by $ N_{\text{gh}} =N_{\theta} + N_{\bar{\theta}}$, then the total degrees of freedom  introduced in the modified LM theory reads 
\begin{equation}\label{eq:DOF_GHOSTS}
     N_{\lambda} - N_{ \text{gh}} + N_{\chi} = 0,
\end{equation}
since $ N_{\lambda} = N_{\chi} = N_{ \text{gh}}/2 =  N_{\phi} $.
Naively, the modified LM theory has a total of
\begin{equation}\label{eq:DOF_M_LM}
        N_{\phi} +N_{\lambda} - N_{ \text{gh}} + N_{\chi}  =  N_{\phi} 
\end{equation}
degrees of freedom, which coincides with the degrees of freedom present in the theory described by the action \eqref{eq:actionPhi} (this is also consistent with the free energy density at finite temperature \cite{lebellac:1996}.) Therefore, the extra degrees of freedom due to the LM field would cancel against the ghost fields of the modified LM theory, while the standard LM theory has twice $ N_{\phi} + N_{\lambda} = 2 N_{\phi} $. 
To count rigorously the degrees of freedom of the modified LM theory 
further investigation is required.

In the next section, we will follow \cite{Brandt:2020gms} to show that the generating functional of the modified LM theory in Eq.~\eqref{eq:GenFuncLM_ModifiedLocal} does not lead to contributions beyond one-loop order, which is a main characteristic of the standard LM theory. Besides that, now that we have introduced ghost fields in the standard LM theory, we show that the doubling of one-loop contributions is absent.

\section{Modified LM theory}
\label{section:modLM}

To obtain a field redefinition invariant path integral for the LM theory we introduce the factor $ \Delta [\phi] $, defined in Eq.~\eqref{eq:def:Delta}, in the measure of integration of the standard LM theory. 
In this section, we shall treat it in more detail showing not only that the additional term $\Delta [ \phi] $ does not lead to contributions beyond one-loop order, but also that the ghost contributions from Eq.~\eqref{eq:DeltaInTermsOfFields} cancel the extra contribution due to the LM field $ \lambda $. 

The generating functional of the modified LM theory, for the action $ S [ \phi ]$ in Eq.~\eqref{eq:actionPhi}, is
\begin{equation}\label{eq:def:fullPathModLM}
    \begin{split}
        \mathcal{Z}_{\text{LM}} [0] ={}&
          \int \mathop{\mathcal{D} \phi} \mathop{\mathcal{D} \lambda} \mathop{\mathcal{D} \bar{\theta} } \mathop{\mathcal{D} \theta} \mathop{\mathcal{D} \chi}  \exp{ \frac{i}{\hbar}  }
         \\ & \times {\int \mathop{d x} \left ( {\mathcal{L}}_{\text{}}^{}  ( \phi ) + \lambda \frac{\delta S[ \phi ] }{\delta \phi} + \bar{\theta} \frac{\delta^{2} S[ \phi ] }{\delta \phi \delta \phi}\theta 
      + \frac{1}{2} \chi \frac{\delta^{2} S[\phi]}{\delta \phi \delta \phi} \chi \right ) }.
    \end{split}
\end{equation}
It can be written conveniently (integrating out the fields $ \theta$, $ \bar{\theta} $, $ \chi $ and $ \lambda $) as
\begin{equation}\label{eq:def:PathIntLM}
        \mathcal{Z}_{\text{LM}}[0] =  \int \mathop{\mathcal{D} \phi} \det \left( \mathcal{L}^{\prime\prime}( \phi )\right)\det \left( \mathcal{L}^{\prime\prime}( \phi )\right)^{-1/2} 
        \mathop{\delta} \left ( \mathcal{L}^{\prime} ( \phi )\right )
    \exp \frac{i}{\hbar} \int \mathop{d x} \mathcal{L} ( \phi ). 
\end{equation}

Using the functional analog of the Eq.~\eqref{A.16}, we obtain
\begin{equation}\label{A.17c}
    \mathcal{Z}_{\text{LM}} [0]= \sum_{\bar{\phi}(x)} \det\left( \mathcal{L}^{\prime\prime}(\bar{\phi})\right)^{-1/2}\exp \frac{i}{ \hbar }  
    \int \mathop{d x}   \mathcal{L}(\bar{\phi}).
\end{equation}
The generating functional of the modified LM theory, at one-loop, now coincides with Eq.~\eqref{eq:ApproximationGenFunPhi}.
The field redefinition invariant formulation of the LM theory leads to the same tree-level and one-loop contributions obtained with the generating functional in Eq.~\eqref{eq:GenFunc}, while higher-loop order contributions vanish. 
It shows that our choice in Eq.~\eqref{eq:def:Delta} led to a modified path integral for the LM theory (in Eq.~\eqref{eq:def:fullPathModLM}) that is invariant under field redefinitions, but that kept restricting quantum corrections to one-loop order.

From Eqs.~\eqref{eq:FactorOf2a} and \eqref{A.17c}, we have that 
 \begin{equation}\label{eq:FactorOf1}
     \begin{split}
         \mathcal{W}_{\text{LM}}  |_{\text{1loop}} &\equiv  - i \hbar \ln \mathcal{Z}|_{\text{1loop}} \\ &= - i \hbar \ln Z|_{\text{1loop}}  =   W |_{\text{1loop}}.
 \end{split} 
\end{equation}
Comparing with Eq.~\eqref{eq:FactorOf2} we see that the factor of $2$ that appears in the standard LM theory is now absent. Hence, considering an appropriate, invariant under field redefinitions, path integral for the LM theory we remove the doubling of one-loop contributions while preserving its main feature, namely the truncated perturbative expansion.

\subsection{Diagrammatic analysis}
Here we give a diagrammatic analysis of the modified LM theory to illustrate the results obtained above. We show that the ghost contributions are responsible for canceling the extra contributions due to the LM field and, as in the Standard LM theory, the perturbative expansion is restricted to one-loop order. 

In the modified LM theory, we have additional terms due to the ghosts $ \theta $, $ \bar{\theta} $, and $ \chi $. Substituting Eq.~\eqref{eq:expandedAction} in Eq.~\eqref{eq:def:Sgh} one reads from the ghost action that 
\begin{equation}\label{eq:Lgh}
    \begin{split}
        \mathcal{L}_{\text{gh}} ( \phi ) ={}&    \bar{\theta}_{i} 
    \left ( a^{(2)}_{ij} + a^{(3)}_{ijk} \phi_{k} + \frac{1}{2!} a^{(4)}_{ijkl} \phi_{k} \phi_{l} + \cdots \right ) 
    \theta_{j} \\ 
                                         & + 
                                         \frac{1}{2} \chi_{i} \left ( a^{(2)}_{ij} + a^{(3)}_{ijk} \phi_{k} + \frac{1}{2!} a^{(4)}_{ijkl} \phi_{k} \phi_{l} + \cdots \right ) \chi_{j} . 
\end{split}
\end{equation}
This shows that the propagators of the ghosts are equal to $ i a_{ij} $ coinciding with the mixed propagators. It is important to remark that the LM field $ \lambda $ does not interact with ghosts, otherwise diagrams of order higher than one loop would arise. 
Now we can proceed with the diagrammatic analysis of the modified LM theory.

First, since ghosts can only appear in closed loops the tree-level contributions are kept unaltered. We can proceed to consider loop diagrams. 
Let us review the results derived from the diagrammatic analysis done in the previous section and check if they remain valid in the presence of the ghost fields introduced in the modified LM theory. 

The conclusions are:
\begin{enumerate}[(i)]
    \item One-loop diagrams with the LM field $ \lambda $ in an external leg are not allowed. 

    \item We can only draw one-loop diagrams with $ \phi $ in the external legs. 

    \item Diagrams with more than one loop are not allowed. 
\end{enumerate}

\begin{figure*}[ht!]
    \centering
    \includegraphics[scale=0.8]{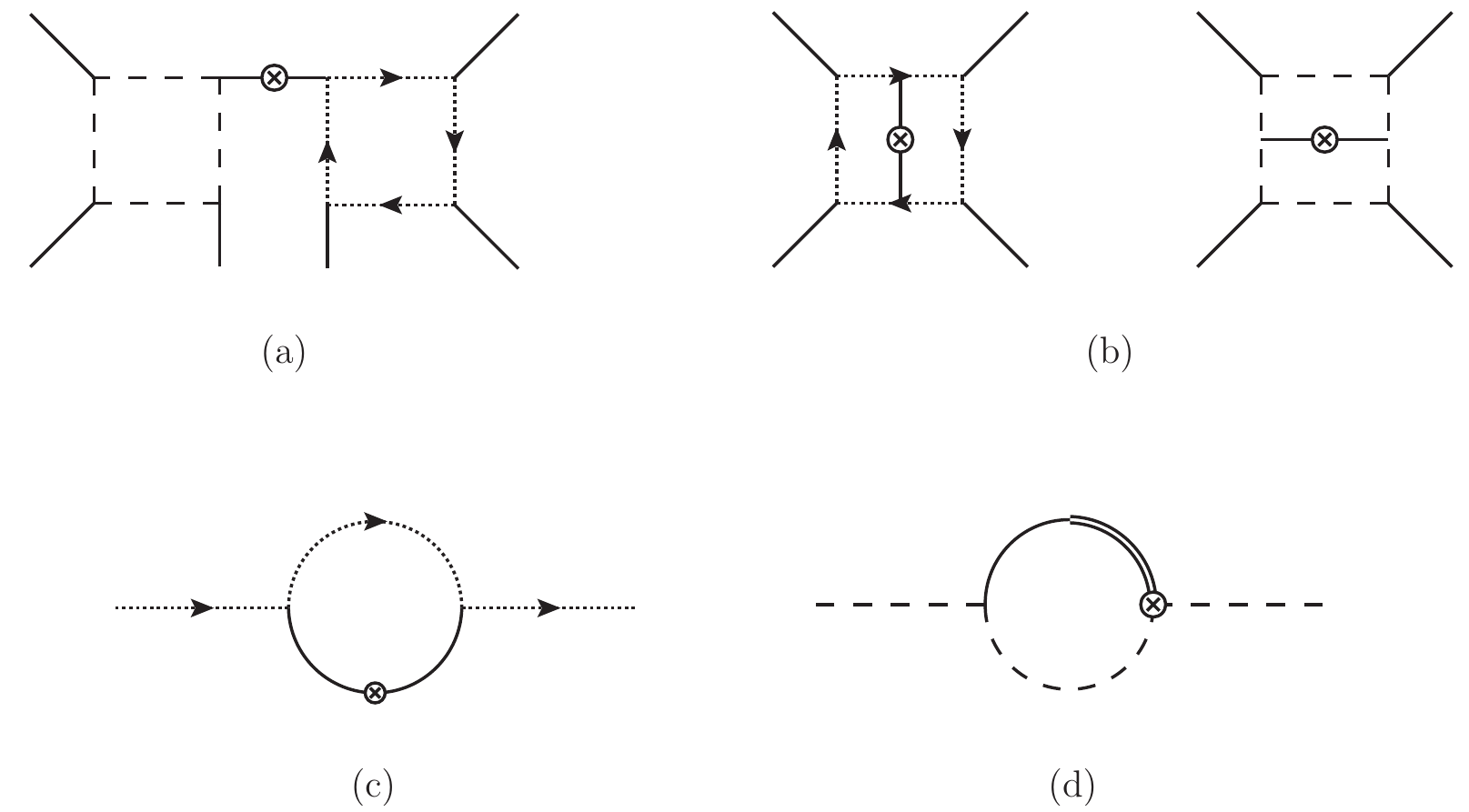}
    \caption{Diagrams higher than one loop order cannot be drawn in the modified LM theory (a, b). Ghosts cannot appear in external lines (c, d). Pointed and dashed lines represent respectively the ghost fields $ \bar{\theta}$, $ \theta $,  and $\chi $ (crosses denote forbidden topologies).}
    \label{fig:6}
\end{figure*}

The ghost fields do not interact with the LM field $ \lambda $, hence the conclusion (i) remains true. It is also not possible to draw any one-loop diagram with external ghosts legs, see Fig.~\ref{fig:6}(d), since we must have at least one internal $ \phi $ propagator, which is forbidden, see Fig.~\ref{fig:6}(c). 
The last conclusion follows from the same reasoning used in section~\ref{section:SLMT}, see Fig.~\ref{fig:6}(a, b). Thus, the ghosts in the modified LM theory do not spoil the main characteristic of the LM theory, namely the restriction of the loop expansion to one-loop order, whereas matter fields require further attention \cite{McKeon:2021qhv}.

\begin{figure*}[ht]
    \centering
    \includegraphics[scale=0.8]{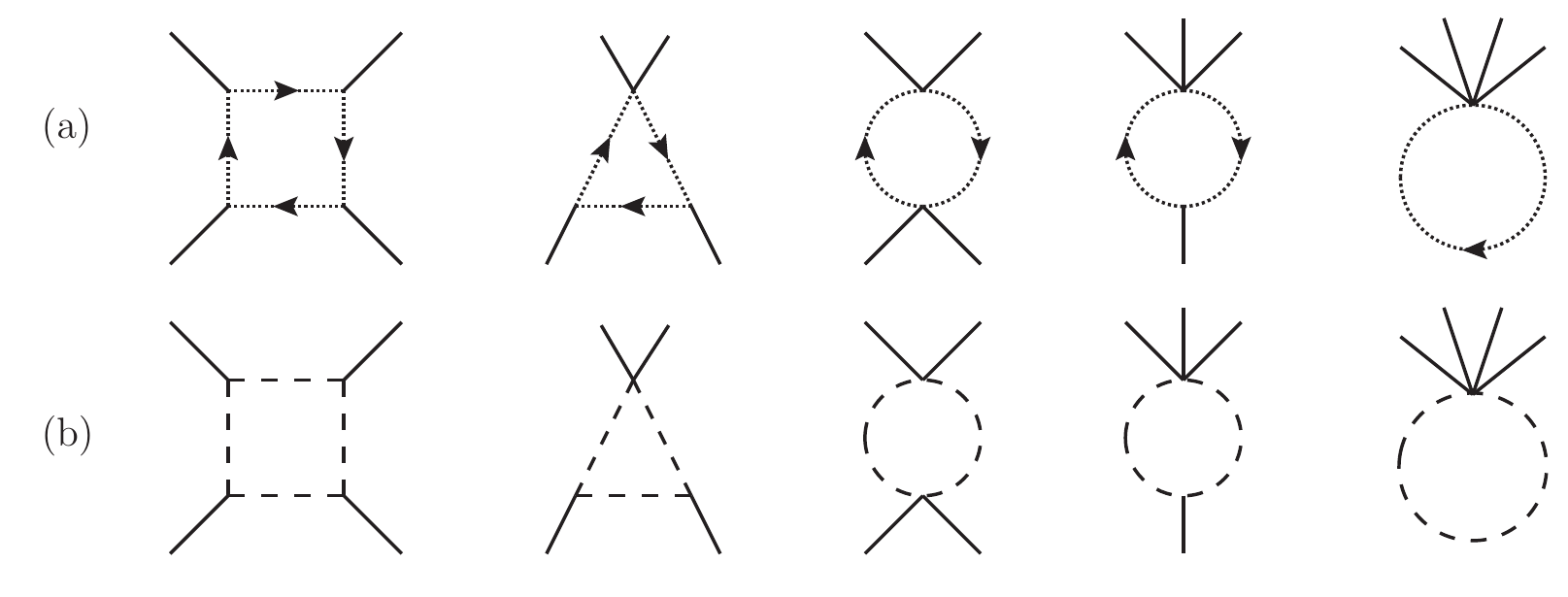}
    \caption{Contributions coming from ghosts to 4-point amplitude $ ( \phi_{i} \phi_{j} \phi_{k} \phi_{l} )$ in the modified LM theory.}
    \label{fig:modLM}
\end{figure*}

To complete our diagrammatic analysis note that the diagrams in the Figs.~\ref{fig:1}, \ref{fig:3}, \ref{fig:4} and \ref{fig:modLM} differ only by overall factors. The diagrams in Fig.~\ref{fig:modLM}(a) have a symmetry factor of $1$ (with the minus sign from the fermionic loop these diagrams have an overall factor of $-1$), while diagrams in Fig.~\ref{fig:modLM}(b) have a symmetry factor of $2$ as the diagrams in Fig.~\ref{fig:3}. Therefore, for these contributions, the ghost diagrams add up leading to an overall factor of $ -1/2$. 

\begin{figure*}[ht!]
    \centering
    \includegraphics[scale=0.8]{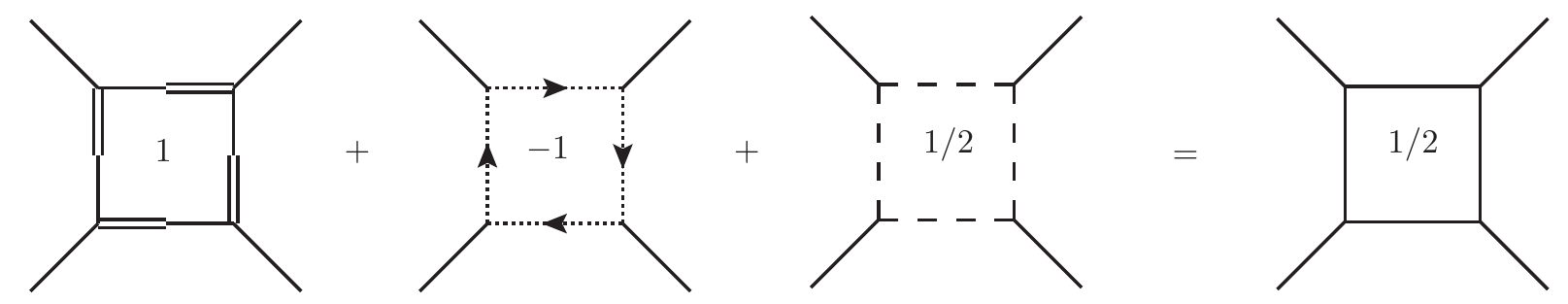}
    \caption{One-loop contribution in the modified LM theory to 4-point amplitude $( \phi_{i} \phi_{j} \phi_{k} \phi_{l} )$, which is identical to the original theory. The overall factors are indicated.}
    \label{fig:NEW}
\end{figure*}

In the modified LM theory the total one-loop contribution is then obtained by summing the diagrams from the standard LM theory in Fig~\ref{fig:1} (overall factor of $1$)  with the ghosts diagrams in Fig~\ref{fig:modLM} (overall factor of $-1/2$) resulting in the usual one-loop contribution (overall factor of $1/2$) coming from the action in \eqref{eq:actionPhi}. For instance, see Fig.~\ref{fig:NEW}.  
This show that ghost contributions in the modified LM theory are canceling the extra one-loop contributions of the standard LM theory, which is consistent with Eq.~\eqref{eq:FactorOf1}.

\section{Discussion}\label{section:discussion}

In this paper, we have found that the path integral of the standard LM theory is not invariant under field redefinitions. To restore this invariance, we proposed a modification of the path integral quantization of the LM theory by introducing a new determinant factor in the measure of the path integral of the standard LM theory. The field redefinition invariant LM formalism turned out to circumvent issues that would otherwise arise in the standard LM theory.
We have shown that, in this modified LM theory, new ghost fields analogous to Lee--Yang ghosts arise. However, these ghosts do not spoil the main feature of the LM theory, namely the restriction of the loop expansion to one-loop order.

Moreover, it is the ghost fields that are responsible for the cancellation of the extra one-loop contributions due to the LM fields, which is one of the main drawbacks of the standard LM theory. We also suggested that ghost fields degrees of freedom should cancel the degrees of freedom coming from the LM fields arguing that the degrees of freedom of the fields with non-vanishing ghost number should be counted as negative. This was justified by the parallel that we have shown between them and Faddeev--Popov ghosts.
But, note that this only holds rigorously for the Faddeev--Popov ghosts since there is a direct correspondence between them and constraints \cite{henneaux1992quantization}.  
This is a shortcoming of our approach that was based solely on path integral quantization. 

A related issue is how to treat singular action in our formalism, in particular the Einstein-Hilbert action.
It can be applied to degenerate theories replacing the original singular action with an effective (non-singular) action obtained through the generalized Faddeev--Senjanovic (FS) procedure \cite{Faddeev:1967fc, Senjanovic:1976br}. This is work in progress, and we expect to report on it in near future; in particular, we want to provide a consistent field redefinition LM theory for quantum gravity, which is the main application of the LM formalism. 

Unitarity is another issue that requires attention in the LM theory.
The standard LM theory is known to yield an unbounded Hamiltonian since the presence of ghost-like LM fields (the kinetic term has a wrong sign) results in instabilities that could lead to the lack of unitarity \cite{McKeon:1992rq, Smilga:2004cy}. In Ref. \cite{McKeon:2021qhv}, this is resolved using indefinite metric quantization. 
In particular, it has been verified that the standard LM theory is consistent with the unitary condition using Cutkosky's cutting rules \cite{Cutkosky:1960sp}. This analysis is not altered by the presence of the ghost fields of the modified LM theory; therefore, both approaches satisfy this unitary condition. 

In our approach the doubling of one-loop quantum correction (and degrees of freedom) are necessarily unphysical and cannot be disregarded. However, we have shown that the ghost fields $ \theta$, $ \bar{\theta} $ and $ \chi $ are responsible for the cancellation of these unphysical one-loop quantum effects coming from the ghost-like LM field $ \lambda $, as Faddeev--Popov ghosts that cancel unphysical longitudinal contributions in gauge theories. Moreover, the unphysical states due to the LM field $ \lambda $ should cancel against the unphysical states coming from the ghost fields $ \theta $, $ \bar{\theta} $, while the third ghost $ \chi $ does not lead to unphysical states as we have discussed in section~\ref{section:FDofQPI}. Therefore, the modified LM theory should be unitary.

To further clarify these issues, the study of the Hamiltonian formalism for the modified LM theory is crucial.  The similarity between the measure \eqref{eq:ModifiedLM_Delta} and a Senjanovic measure \cite{Senjanovic:1976br} 
\begin{equation}\label{eq:measureSenja}
    \mathop{\mathcal{D} p} \mathop{\mathcal{D} q} 
    \left | \det \lVert \left \{\alpha_{i}  ,\, \alpha_{j}  \right\} \rVert \right|^{1/2} \delta ( \alpha_{k}),
\end{equation} 
where $ \alpha_{k} $ are second class constraints, after the canonical momenta $p$ are integrated; can be a clue to investigate it. 
If this is the case, the modified LM theory is a degenerate (higher-derivative) theory. The analogy between Faddeev--Popov ghosts and the ghosts $ \theta $, $ \bar{\theta} $ of the modified LM theory would be confirmed, and we could rigorously state that they count as negative degrees of freedom. Moreover, since the phase space of the LM theory would be reduced (it is constrained), it should be free of Ostrogradsky ghosts \cite{Motohashi:2016ftl, Chen:2012au}, which agrees with our results.

Besides that, it is clear that there is also a similarity between the classical path integral \cite{Gozzi:1986ge} 
  \begin{equation}\label{eq:CPI}
          Z_{\text{CPI}}[0] = 
      \int \mathop{\mathcal{D} \phi} \mathop{\mathcal{D} \lambda} \mathop{\mathcal{D} \bar{\theta} } \mathop{\mathcal{D} \theta} 
      \exp{i \int \mathop{d x} \left ( \lambda \frac{\delta S[ \phi ] }{\delta \phi} + \bar{\theta} \frac{\delta^{2} S[ \phi ] }{\delta \phi \delta \phi}\theta \right ) }
  \end{equation}
  and the path integral of the LM theory, in particular of the modified LM theory in Eq.~\eqref{eq:def:fullPathModLM}.
  The classical path integral in Eq.~\eqref{eq:CPI} restricts the field $ \phi $ to its classical field configurations, while in the LM theory in Eq.~\eqref{eq:def:fullPathModLM} the field $ \phi $ is restricted to satisfy its classical equations of motion. At first sight, these constraints appear to be equivalent, but the classical path integral is invariant under several novel symmetries as the (anti)BRST-like symmetry \cite{Gozzi:1989bf, Gozzi:1986ge}, supersymmetry \cite{Gozzi:2000sf, Deotto:2000ia}, and universal local symmetries \cite{Gozzi:2010iq} that are broken by the weight $ \exp \left ( iS [ \phi ]/\hbar\right ) $ in the quantum path integral in Eq.~\eqref{eq:def:fullPathModLM}.  Therefore, although the similarities between the generating functionals \eqref{eq:CPI} and \eqref{eq:def:fullPathModLM}, they cannot be equivalent. In particular, the properties of non-superposition and non-interference, found in the classical theory, correspond to the universal local symmetries of the classical path integral \cite{Gozzi:2010iq}. Considering that the path integral of the LM theory is not invariant under them, it must contain richer physics that we hope to explore in future work.

\section*{Acknowledgments}
Discussions with J. Frenkel and D. G. C. McKeon were illuminating.
We would also like to thank Pedro F. H. Bairrão for a careful reading of the manuscript and J. P. Edwards for an enlightening comment. 
We thank CNPq (Brazil) for financial support.

\appendix

\section{Modified LM theory with Grassmann fields}\label{section:FSandGofLM}

In the case of fermionic systems described by anticommuting fields (Grassmann fields) the measure of integration changes under field redefinitions in the inverse way of the bosonic case:
\begin{equation}\label{eq:measureintegrationfermions}
    \mathop{\mathcal{D} \psi} \mathop{\mathcal{D} \bar{\psi}} \to \det \bm{J}_{\psi}^{-1}  \mathop{\mathcal{D} \psi'} \mathop{\mathcal{D} \bar{\psi}}, 
\end{equation}
where the Jacobian determinant is defined analogously with the bosonic case (in Eq.~\eqref{eq:JacobianF}) as 
\begin{equation}\label{eq:def:detJpsi}
    \det\bm{J}_{\psi}^{-1} = \det\frac{\partial \psi '}{\partial \psi }. 
\end{equation}

Thus, in the LM formalism we have that 
\begin{equation}\label{eq:FDofMIF}
    \mathop{\mathcal{D} \psi} \mathop{\mathcal{D} \bar{\psi}}     \mathop{\delta} \left ( \frac{\delta S  }{\delta \psi}\right ) \mathop{\delta} \left ( \frac{\delta S }{\delta \bar{\psi}}\right )   \to \det \bm{J}_{\psi}^{-2}  
    \mathop{\mathcal{D} \psi'} \mathop{\mathcal{D} \bar{\psi}}     \mathop{\delta} \left ( \frac{\delta S'}{\delta \psi'}\right ) \mathop{\delta} \left ( \frac{\delta S'}{\delta \bar{\psi}}\right ).
\end{equation}
To restore the usual field redefinition property, as in Eq.~\eqref{eq:measureintegrationfermions}, we introduce in the measure the term $ \Delta [ \psi , \bar{\psi} ]$ that must transform as 
\begin{equation}\label{eq:TofDelta}
    \Delta [ \psi , \bar{\psi} ] \to 
 \Delta'[ \psi', \bar{\psi} ]=
    \det \bm{J}_{\psi} \Delta [ \psi' , \bar{\psi} ].
\end{equation}

We can easily generalize Eq.~\eqref{eq:def:Delta} to a fermionic system as 
\begin{equation}\label{eq:def:DeltaFermions}
    \Delta [ \psi , \bar{\psi} ] = \det \left(\frac{\delta^{2} S }{\delta \psi\delta \bar{\psi}}\right)^{-1}.
\end{equation}
It satisfies Eq.~\eqref{eq:TofDelta}, since Eq.~\eqref{eq:transDerivative} also is valid for left derivatives: 
\begin{equation}\label{eq:TofLDerivative}
    \frac{\delta }{\delta \psi} = \frac{\delta \psi ' }{\delta \psi} \frac{\delta }{\delta \psi '} = \bm{J}_{\psi}^{-1} \frac{\delta }{\delta \psi '}, 
\end{equation}
therefore,
\begin{equation}\label{eq:TofDeltaF}
    \Delta ' [ \psi ', \bar{\psi} ] = \det \left ( \bm{J}_{\psi}^{-1} \frac{\delta^{2} S }{\delta \psi ' \delta \bar{\psi}} \right )^{-1}= \det \bm{J}_{\psi} \Delta [ \psi ' , \bar{\psi} ]. 
\end{equation}

As in the bosonic case, the contribution of $ \Delta[ \psi , \bar{\psi} ]$ cancels half of the one-loop contribution from the standard LM formalism. For instance, let us consider a free fermionic system. The one-loop contribution is equal to $  \det M = \Delta_{0}^{-1} =\det \left ( \delta^{2} S_0/\delta \psi \delta \bar{\psi}\right ) $, in the LM formalism we have the square of it: $ \det M^{2} $ (the doubling). Instead, in the modified LM theory, we obtain $ \Delta_0 \det M^{2}= \det M^{-1} \det M^{2} = \det M $ as in the original theory. Moreover, the contributions beyond one-loop order are eliminated as in the standard LM formalism. 

In this fermionic system, the measure of integration in the framework of the modified LM formalism is obtained by the substitution
\begin{equation}\label{eq:MofFSwLM}
    \mathop{\mathcal{D} \psi} \mathop{\mathcal{D} \bar{\psi}}
    \xmapsto[\text{theory}]{\text{LM}}  
    \mathop{\mathcal{D} \psi} \mathop{\mathcal{D} \bar{\psi}}      
    \det \left ( \frac{\delta^{2} S[ \psi , \bar{\psi} ] }{\delta \psi \delta \bar{\psi}}\right )^{-1} 
    \delta \left ( \frac{\delta S  }{\delta \psi}\right ) \mathop{\delta} \left ( \frac{\delta S }{\delta \bar{\psi}}\right ).
\end{equation}
This is mostly interesting for gauge theories since Faddeev--Popov ghosts are Grassmann fields. 

The Eqs.~\eqref{eq:MofFSwLM} and \eqref{eq:ModifiedLM_DeltaTransforms} can be written in a single expression using the concept of superfields. Let $ \Psi_{i} $ denote a set of fields with even and odd Grassmann numbers, bosonic and fermionic fields respectively, that define the superfield $ \Psi $. The measure of integration of the generating functional in the modified LM theory for the superfield $ \Psi $ reads
\begin{equation}\label{eq:generalLMmeasure}
    \mathop{\mathcal{D} \Psi} 
    \xmapsto[\text{theory}]{\text{LM}}  
    \mathop{\mathcal{D} \Psi} \mathop{\mathrm{SDet}} \left ( \frac{\delta^{2} S[ \Psi ] }{\delta \Psi \delta \Psi}\right )^{+1/2} \mathop{\delta} \left ( \frac{\delta S [ \Psi ] }{\delta \Psi}\right ),
\end{equation}
where $ \mathop{\mathrm{SDet}} M $ denote the superdeterminant of the supermatrix $M$.

\section{Additional symmetries in the modified LM theory}\label{section:symmetries}

\subsection{Ghost number symmetry}

The effective action of the modified LM theory in Eq.~\eqref{eq:def:Seff} is invariant under the transformation 
\begin{equation}\label{eq:transform:ghost}
    \delta \bar{\theta} = - \sigma \bar{\theta} , \quad \delta \theta = \sigma \theta ,  \quad  \delta \phi = \delta \lambda = \delta \chi =0;
\end{equation}
where $ \sigma $ is a commuting parameter. 

This symmetry of the LM theory is related to the conservation of the ghost charge $Q_{\text{gh}}$ \cite{Gozzi:1986ge}. The ghost charge is directly related to the ghost number $ \mathop{gh}$ mentioned in section~\ref{section:FDofQPI}. It can be shown that $ Q_{\text{gh}} \theta = \sigma $, $ Q_{\text{gh}} \bar{\theta}=- \sigma $ while $ Q_{\text{gh}} \phi = Q_{\text{gh}} \lambda = Q_{\text{gh}} \chi =0$, that is, $ Q_{\text{gh}} \equiv \sigma \mathop{gh} $.

\subsection{(Anti)BRST-like symmetry}

The ghost sector in Eq.~\eqref{eq:def:Seff} is invariant under the BRST-like symmetry
\begin{equation}\label{eq:transform:ghostS}
    \delta \theta =  \chi \epsilon, \quad 
    \delta \chi = \epsilon \bar{\theta},
    \quad \delta \phi = \delta \lambda = \delta \bar{\theta} =0,
\end{equation}
and the anti-BRST symmetry
\begin{equation}\label{eq:transform:ghostS2}
    \bar{\delta} \bar{\theta} =   \chi \bar{\epsilon}, \quad \bar{\delta} \chi = \theta \bar{\epsilon},
     \quad \bar{\delta} \phi = \bar{\delta} \lambda = \bar{\delta} \theta =0,
\end{equation}
where ($ \bar{\epsilon} $) $ \epsilon $ is an anticommuting parameter of the (anti)BRST symmetry.

The idempotency of the BRST operator can be shown straightforwardly: 
\begin{equation}\label{eq:IPofBRST}
    \delta^{2} \chi = \delta \left ( \epsilon \bar{\theta}\right ) = \epsilon \delta \bar{\theta} =0
\end{equation}
and 
\begin{equation}\label{eq:IpofBRST2}
    \delta^{2} \theta = \delta \left ( \chi \epsilon\right ) =- \epsilon^{2} \bar{\theta}=0.
\end{equation}

\bibliography{ghosts.bib}      

\end{document}